# Solar modulation of cosmic ray positrons in a very quiet heliosphere

**Marius S Potgieter[1], Etienne E Vos, Driaan Bisschoff, Jan-Louis Raath**
*Centre for Space Research, North-West University, 2520 Potchefstroom, South Africa*
*E-mail:* `marius.potgieter@nwu.ac.za`

**Mirko Boezio, Riccardo Munini**
*INFN, Sezione di Trieste I-34149 Trieste, Italy*

**Valeria Di Felice**
*INFN, Sezione di Rome "Tor Vergata", I-00133 Rome, Italy.*
*Also: Agenzia Spaziale Italiana (ASI) Science Data Center, I-00044 Frascati, Italy*

Since the beginning of the space exploration era, solar activity was observed at its lowest level during 2006 to 2009. During this period, the PAMELA space experiment observed spectra for galactic cosmic rays, specifically for protons, electrons and positrons over a wide energy range, during what is called an $A < 0$ solar magnetic polarity cycle. Drift theory predicts a difference in the behaviour for these oppositely charge particles during $A < 0$ cycles. An opportunity was thus created to study the predicted charge-sign-dependent modulation, also now for very quiet heliospheric conditions. A comprehensive three-dimensional, drift modulation model has been used to study the solar modulation for cosmic rays in detail with extensive comparison to the observed PAMELA spectra for the mentioned period. First, this was done for protons and secondly for electrons, as already published, to test and to authenticate the modelling approach and then to come to a better understanding and appreciation of the underlying physics, such as diffusion and drift theory. The results were also used to make predictions of how cosmic rays are differently modulated down to low energies (1 MeV) for the two magnetic polarity cycles of the Sun, and what role drifts play in this process. All computed solutions are based on new very local interstellar spectra, now also done for positrons. This report is focussed on detailed aspects of the solar modulation of positrons during the extraordinary quiet solar modulation period from 2006 to 2009. For the first time, a meaningful modulation factor in the heliosphere is computed for positrons, from 50 GeV down to 1 MeV, as well as the electron to positron ratios as a function of time and rigidity for the mentioned period.

---

[1]Speaker

# 1.  Introduction

The theory for gradient and curvature drifts of cosmic rays (CRs) in the heliosphere predicts that the modulation of positively charged CRs should develop differently compared to negatively charged CRs. For example, during $A < 0$ cycles, the heliospheric magnetic field (HMF) is directed outward in the northern and inward in the southern hemisphere, causing protons and positrons to drift from the heliospheric boundary inwards to the Earth mainly through the equatorial regions of the heliosphere, while electrons and antiprotons drift inwards mainly through the polar regions. Protons and positrons then encounter the changing, wavy heliospheric current sheet (HCS) so that their recovery to solar minimum modulation depends to a large extent on the decreasing rate of the tilt angle of the HCS [1]. The intensity-time profile of protons before and after solar minimum modulation exhibits a peak during $A < 0$ magnetic polarity cycles, whereas electrons being less sensitive to what happens to the tilt angle, typically exhibit a profile that is less peaked. If particle drifts would completely dominate the modulation process, the profiles for protons and positrons would be sharply peaked while the profiles for electrons and anti-protons would be flattish. Of course, when the HMF changes its polarity to an $A > 0$ cycle, protons and positrons will exhibit the flatter profiles while electrons and antiprotons exhibit the sharper profiles, depicting a clear 22-year cycle. This predicted 22-year periodicity has been observed with CR detectors on ground level (called neutron monitors) since the late 1950s. This mainly constitutes what is known as charge-sign-dependent solar modulation caused by particles drifts; see e.g. reviews [2,3].

Precise and continuous measurements of positron and antiproton spectra over a wide energy range, done with the same instrument, will contribute significantly to the understanding of charge-sign-dependent modulation, especially if these observations are taken from one solar minimum to another, thus including the period of maximum solar activity and the reversal of the HMF polarity. Fortunately, this has become a reality with the PAMELA [e.g. 4] and AMS02 [e.g. 5] space missions. PAMELA has provided a perfect opportunity to study charge-sign-dependent modulation during an $A < 0$ minimum modulation cycle, from June 2006 to the end of 2009, reporting differential fluxes for protons [6], electrons [7] and positrons [8,9]. And, recently, also reported the electron to positron ratio for over 9 years including the polarity reversal period in 2013-14 [10].

The PAMELA data prompted [11-14] to study the modulation of galactic protons and electrons using a comprehensive three-dimensional (3D) modulation model that includes drifts and all the other important solar-heliospheric modulation processes. This has been done elaborately for protons and electrons. The observed different behaviour between protons and electrons for the mentioned solar minimum period was extensively reported also by [15]. However, this type of study has not yet been done extensively for positrons, and is the focus of this current work. The PAMELA positron data from [9] is used, which is still considered as preliminary. We compliment these positron observations for the period 2006 to 2009 with the solutions from the 3D numerical modelling by following the modulation approach as reported by [11,14].  Previous and older modeling on positrons in particular was reported by [16].

It is by now well-known that heliospheric modulation conditions were exceptionally quiet during the mentioned time period. This was discussed in detail by [12]. The current study will also contribute to the evaluation of long-term positron observations, the very local interstellar spectrum for positrons and when compared to numerical models, the impact it may have on underlying theories, in particular drift theory. But, at first only the general features of positron modulation during a very quiet heliosphere is shown.

## 2.  The very local interstellar spectrum for cosmic ray positrons

Computational modelling of CR modulation requires that local interstellar spectra be specified as input to models for the respective CR particles, such as protons, electrons, helium, oxygen, and more [e.g. 17]. This is done at the boundary of the simulated heliosphere, assumed to be the heliopause (HP). These very local interstellar spectra (VLIS) for protons and electrons were discussed in detail by [11-14]. Numerical studies also produce a set of modulation parameters for protons and electrons that can be considered as appropriate and applicable to other CR species. The entire parameter set for electron modulation, specifically the result that the diffusion coefficients (DCs) for electrons is almost independent of rigidity $R$ (or kinetic energy $E$) between ~ 5 MV and ~ 0.4 GV [11,15], is assumed also applicable to positron modulation. This means that for these two oppositely charged CRs the only cause of difference in modulation is assumed to be drifts. If the VLIS for electrons and positrons were known exactly, drift theory could have been tested rigorously, but unfortunately this is not the case. So the approach we follow is to take drift theory as it stands at present, and by comparing model solutions for electrons and positrons with observations, we test if the positron LIS from GALPROP, the well-known Galactic propagation code [e.g. 18; available on-line], is suitable for modulation studies, and if not, find an improved VLIS for positrons.

Following this approach, we report a modified VLIS for positrons as shown in Fig. 1, together with observed positron spectra from PAMELA (combined 2006 to 2009) and AMS02 (combined 2011 and 2013) as cited in the figure's caption. Because the focus of this study is on solar modulation, the reported observed positron excess evident above ~10 GeV, is seen as an astrophysical effect and does not influence our modulation results [19]. The solid line is the VLIS for positrons from this study; the two other LIS's (dashed and dotted grey lines) are obtained by applying GALPROP following two modelled approaches known as 'plain diffusion' (DP) and 'plain diffusion with reacceleration' (PDR) in Galactic space. It follows that below ~200 MeV, the PDR approach is similar to what we find, but that the PD approach is too low, a deviation already starting below ~ 2 GeV. But, between 200 MeV and 2 GeV the PDR approach is far too high. If each of the two GALPROP approaches was used without modification, the PAMELA positron spectrum at the Earth could not be reproduced. The VLIS given by the solid black line and used in the numerical modulation model does so, as is illustrated next.

## 3.  The numerical model

The 3D model is described in detail by [11-14,20]. Briefly, it is based on using the VLIS for positrons as given above in an assumed heliosphere with the HP at 122 AU. It contains a heliosheath between the termination shock (TS) and HP but neglects the reaccelerated effects of the TS [16]. The VLIS's are modulated up to the Earth. In order to do so, the three DCs and the drift coefficient must be specified, in particular their $R$ dependence. It suffices to say that the DCs are independent of $R$ below ~ 0.4 GeV for both electrons and positrons as motivated by [11-14]. This means that the spectral shape of the modulated positron spectra below ~ 100 MeV resembles that of the VLIS; at larger radial distance even more so towards higher energies as adiabatic energy loses become progressively less than in the inner heliosphere. The three DCs and the drift coefficient increase with time proportional to $1/B$, with $B$ the observed magnitude of the HMF at the Earth. In additional the tilt angle $\alpha$ of the wavy HCS decreases from 2006 to 2009. For an illustration of these assumptions and effects, see Fig. 4 in [12].

## 4. Comparison of modeling results with observations

In Fig. 2, the PAMELA positron spectrum for the second half of 2006, indicated as 2006b, is shown in comparison with computed spectra at the Earth (1 AU) and at radial distances of 10, 50 and 100 AU, with respect to the new VLIS specified at 122 AU. Details are given in the figure's caption. These computations are repeated for comparison with PAMELA spectra for the periods 2007a,b, 2008a,d, and 2009a,b, and shown in Fig. 3. Apart from changing the tilt angle and the magnitude of the HMF from 2006b to 2009b, the DCs had to be increased additionally at lower energies by a factor of 1.5 for this period; see also results by [11-15,20].

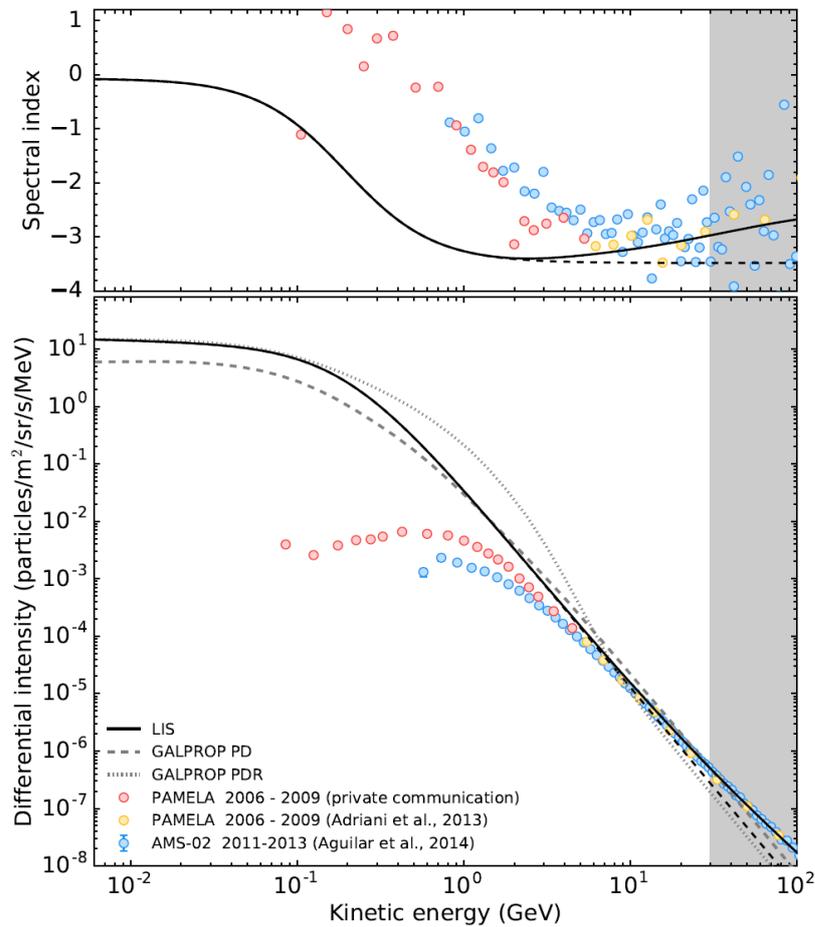

**Figure 1:** A newly constructed positron VLIS (solid black line), with preliminary positron measurements from PAMELA [9] below 5 GeV (red symbols), above 5 GeV [8] (orange symbols), and AMS02 data [5] (blue symbols) as indicated. Top panel gives the spectral index for these measurements and the constructed VLIS. Bottom panel shows various energy spectra: dashed black lines for $E > \sim 2$ GeV serves as $E^{-3.5}$ reference line; dashed and dotted grey lines represent the 'plain diffusion' (PD) and 'plain diffusion with reacceleration' (PDR) propagation regimes in the Galaxy as used within GALPROP [e.g. 18], respectively.

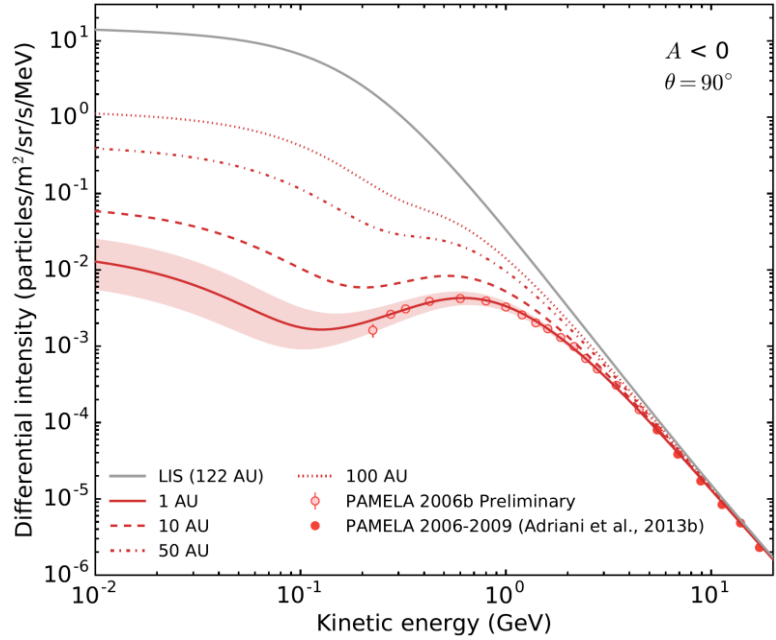

**Figure 2:** Computed positron spectra using DCs from [11] at radial distances of 1 AU, 10 AU, 50 AU and 100 AU, with respect to the new VLIS specified at 122 AU, for an A < 0 cycle in the equatorial plane. Filled symbols are PAMELA observations [8] for 2006b (second half of 2006). The open symbols, down to ~200 MeV, are regarded as work in progress and subject to possible changes [9]. Positron observations down to 80 MeV are still evaluated and not shown here. Red shaded band is considered as the upper and lower limits of what the model predicts at lower energies based on the spread in the observations at higher energies.

In Fig. 4, the corresponding modulation factor (MF) for positrons from 50 GeV down to 1 MeV, calculated by taking the ratio of the VLIS at 122 AU to the modulated spectra at the Earth, as shown in Fig. 3. The smaller this factor is, the more modulation takes place. Grey shaded bands indicate MF regimes e.g. above 15 GeV, the MF is > 0.9 and so on. It is interesting to note that the largest MF occurs between 80-100 MeV for positrons, also evident from Fig. 3, and is the result of incorporating drift effects in the model; see also [15; references there-in].

### 4.1 Electron to positron ratio ($e^-/e^+$)

In what follows, the electron modelling computations published by [11,13,15,20] are used together with the positron computations as shown above in Fig. 3, to compute $e^-/e^+$ ratios as a function of time for the extra-ordinary quiet minimum period between 2006 and 2009. This is shown in Fig. 5, for different ranges of rigidity *R*, e.g., between 10 MV and 30 MV (dark-blue lines); between 120 MV and 70 GV (light-blue to red lines), according to the colour bar shown on the right. Ratios are normalized with respect to 2006b.

It should be noted that below about 50 MeV electron intensities (spectra) are significantly changed at the Earth because of the contribution from Jovian electrons which is not included here; see e.g. [21].

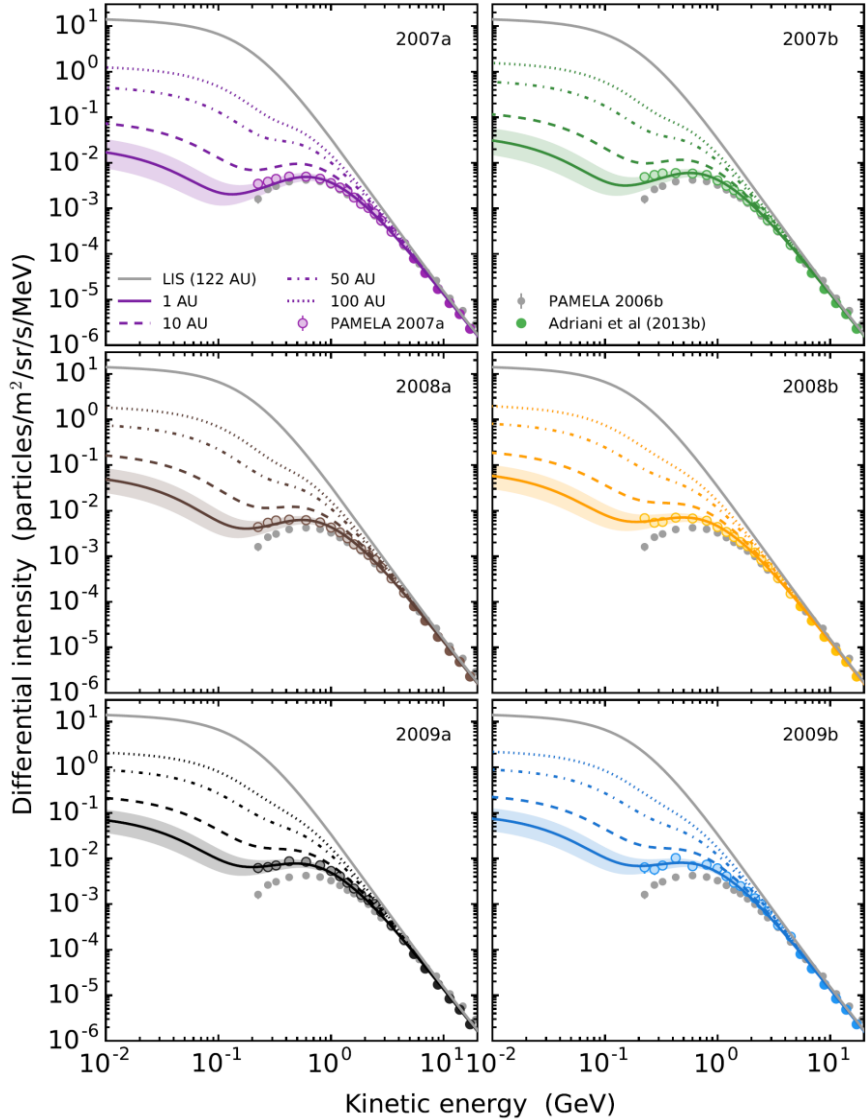

**Figure 3:** Similar to Fig. 2, now also for computed and measured positron spectra for the periods 2007a to 2009b (first half of 2007 to second half of 2009) with respect to 2006b (grey data points). The coloured, shaded spectral bands have the same meaning as in Fig. 2.

## 5.  Discussion and Conclusions

The solar modulation of positrons is studied in detail with a comprehensive 3D drift model in comparison with positron observations from the PAMELA space experiment for the period 2006 to 2009 which is acknowledged as a period of extra-ordinarily low solar activity. This follows on similar studies of cosmic ray protons and electrons. The latter was used to establish the spatial and rigidity dependence of the three diffusion coefficients as well as for the drift coefficient, thus finding a full set of modulation parameters assumed to be fully applicable also to positrons in the heliosphere. It is found that when this same set of modulation parameters and heliospheric features are used for positrons, the PAMELA positron observations cannot be reproduced as good as for electrons. The logical option was to modify the positron VLIS as computed with a single propagation model in the GALPROP code.

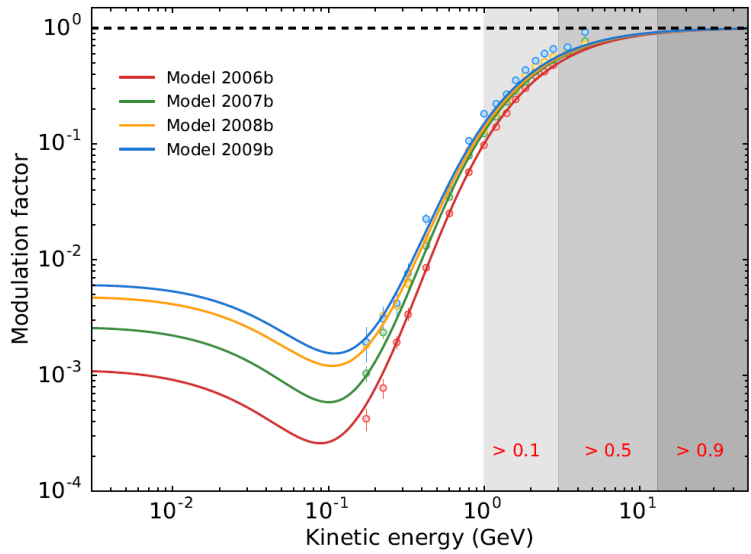

**Figure 4:** The modulation factor (MF) for positrons from 50 GeV down to 1 MeV, calculated by taking the ratio of the VLIS at 122 AU to the modulated spectra at the Earth (1 AU). Solid lines and symbols represent the model solutions and PAMELA observations, respectively, for the second halves of each year, from 2006-2009. Grey shaded bands emphasize regimes of the MF.

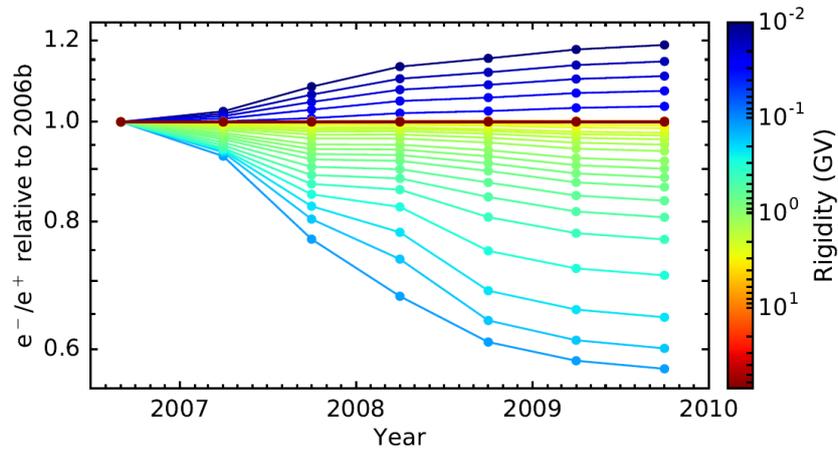

**Figure 5:** Modeled (computed) $e^-/e^+$ ratio as a function of time, 2006 to 2009, for different rigidity ranges, according to the colour bar on the right; e.g. between 10 MV and 30 MV (dark-blue lines); between 120 MV and 70 GV (light-blue to red lines). All ratios are normalized with respect to 2006b. Note that for high rigidities, reddish lines, the ratio remains unchanged.

This new positron VLIS was then used to conduct this study with remarkably good agreement between the model and the observations in terms of time and energy. For the first time, a meaningful modulation factor is computed for positrons, from 50 GeV down to 1 MeV, as well as the $e^-/e^+$ ratios as a function of time and rigidity for the period 2006 to 2009.

The South African authors acknowledge the partial financial support of the South African National Research Foundation (NRF). The Italian authors acknowledge the partial financial support from the Italian Space Agency (ASI) under "Programma PAMELA - attività scientifica di analisi dati in fase E".